\documentclass[11pt]{article}
\usepackage{amsmath}
\usepackage{amsfonts}
\usepackage{amssymb}
\usepackage{epsfig}
\usepackage{times}
\newcommand{\be}{\begin{equation}}
\newcommand{\ee}{\end{equation}}
\newcommand{\ba}{\begin{array}}
\newcommand{\ea}{\end{array}}
\newcommand{\bea}{\begin{eqnarray}}
\newcommand{\eea}{\end{eqnarray}}

\def\ket{\,\right\rangle}
\def\bra{\left\langle\, }

\def\d{\rm d}

\def\d{\rm d}

\def\hbar{\not{\hbox{\kern-2.3pt $h$}}}
\def\psl{\not{\hbox{\kern-2.3pt $p$}}}
\def\Psl{\not{\hbox{\kern-2.3pt $P$}}}
\def\ksl{\not{\hbox{\kern-2.3pt $k$}}}
\def\qsl{\not{\hbox{\kern-2.3pt $q$}}}
\def\slad{\not{\hbox{\kern-2.3pt $\partial$}}}
\def\I{\rm i}
\def\Esl{\not{\hbox{\kern-2.3pt $\epsilon$}}}
\begin{document}
\begin{titlepage}
%

\hfill PAR-LPTHE 01-54
\vskip 4.5cm
{\baselineskip 17pt
\begin{center}
{\bf ON THE $\;$ B$ \to$ J$/{\bf \Psi}$ + K$^*$$\;$ DECAY}
\end{center}
}

\vskip .5cm
\centerline{{
\em      Xuan Son Nguyen 
\footnote[1]{E-mail: xuanson@lpthe.jussieu.fr}
 and  Xuan-Yem  Pham}
\footnote[2]{E-mail: pham@lpthe.jussieu.fr} }

\vskip 3mm
\vskip 2mm
\centerline{{
\em Laboratoire de Physique Th\'eorique et Hautes Energies, Paris}
     \footnote[3]{LPTHE tour 16\,/\,1$^{er}\!$ \'etage,
          Universit\'e P. et M. Curie, BP 126, 4 place Jussieu,
          F-75252 PARIS CEDEX 05 (France).}
}
\centerline{\em Universit\'es Paris 6 et Paris 7;} 
\centerline{\em Unit\'e associ\'ee au CNRS, UMR 7589.}
\vskip 1.5cm

{\bf Abstract}

Within the QCD factorization approach, we calculate the {\it process}- and {\it polarization}-dependent 
nonfactorizable terms $\widetilde{a}_{\lambda}$ of the B$\to$ J$/\Psi +{\rm K^*}$ decay. The longitudinal part
$\,\widetilde{a}_{0}\,$ is infrared convergent and large enough to agree with recent experimental data, provided that the B-K$^*$ form factors $A_1(m^2_{\Psi})$ and $A_2(m^2_{\Psi})$ satisfy some constraints
 met by some (but not all) models.
  The transverse parts $\widetilde{a}_{\pm}$ on the other hand are infrared divergent, 
the procedure used to handle such divergence  is discussed in relation with the B$\to$ J$/\Psi +{\rm K}$ case in which the same problem arises.
Our nonzero phases of the helicity amplitudes are consistent with experimental data recently measured for the first time by the CDF and BaBar groups.  
\smallskip

{\bf PACS} number(s): 13.25.Hw, \quad 12.38.Bx, \quad 12.39.Hg 
\vfill
\end{titlepage}
%
%
\section{Introduction}
\label{section:introduction}
Among a hundred hadronic decay modes of the B mesons\cite{data}  investigated from both experimental and theoretical sides,
 the process  B$ \to {\rm J}/\Psi +{\rm K}^*(892)$ is particularly interesting for many reasons:

(i) - It  is the first color-suppressed B decay observed in 1994 by the Argus group\cite{Argus} with the largest branching ratio for its  class. Since then, important measurements are intensively explored in great detail by the Cleo\cite{Cleo}, CDF\cite{CDF}, BaBar\cite{Babar} and Belle\cite{Belle} collaborations.  

Using both angular and time distributions to separate the CP-even from the CP-odd eigenstates, the asymmetry between B$ \to {\rm J}/\Psi +{\rm K}^*$ and $\overline{\rm B} \to  {\rm J}/\Psi +\overline{\rm K}^*$ -- which directly  gives
 $\sin 2\beta$  at the same degree of precision as its companion "golden" mode 
B$(\overline{\rm B}) \to {\rm J}/\Psi +{\rm K}_{\rm S}$ --  is central to our understanding of the standard Kobayashi-Maskawa CP violation mechanism. 
Furthermore, new physics beyond the Standard Model
can be hinted by  consistently comparing the $\beta$ angle obtained from this CP asymmetry with its unitarity triangle value determined by other experiments (the $\epsilon$ measurement in K decays, B$^0$-$\overline{\rm B}^0$ mixing, $V_{ub}/V_{cb} ...$). Are they unambigously equal ? 

(ii) - Over both the vector + pseudoscalar
B$\to$ V + P and the pseudoscalar + pseudoscalar B$\to$ P + P modes, the advantage of the vector + vector B$\to$ V + V decays (B$ \to {\rm J}/\Psi +{\rm K}^*$ considered
 here) stands out in the possibility of detailed analyses of the three helicity amplitudes.  The three decay amplitudes (one longitudinal and two transverse) denoted by H$_{\,0}$, H$_{-1}$ and H$_{+1}$ could be separately determined both for their  magnitudes $|{\rm H}_{\lambda}|$ and phases $\delta_{\lambda}$. These $|{\rm H}_{\lambda}|$ and
$\delta_{\lambda}$ analyses provide a powerful tool to test not only the naive factorization method\cite{Gourdin} -- usually adopted  to deal with exclusive two-body hadronic decays -- but also the real and imaginary part of the nonfactorizable terms  which are calculable in QCD approaches\cite{Beneke, Keum} can then be confronted with experiments.
We note that  such analysis cannot be done in B$\to$ V+P and B$\to$ P+P 
decays since only the absolute value of a single amplitude (the equivalent of  
$|{\rm H}_{0}|$) can be measured in these processes.

(iii) - The transverse H$_{-1}$ and H$_{+1}$  amplitudes with both magnitudes and phases  provide also an useful
 way to test robustness of factorization manifested through the $V\mp A$ property of the effective weak currents in the Wilson operator product expansion (OPE). It would answer
the question\cite{Suzuki} whether or not the 
 interactions between the hadronic decay products, usually called final-state 
interactions (FSI), are strong enough to flip the quark spin in 
color-suppressed B decays. Although intuitively this spin flip 
unlikely occurs, this possibility could be tested however.

(iv) - Improved  by QCD which gives the $\alpha_{s}$ corrections to the decay amplitudes in OPE, the three "heavy to light" B-K$^*$ form-factors in B$\to {\rm J}/\Psi + {\rm K}^*$  : 
A$_1(q^2)$,  A$_2(q^2)$ and V$(q^2)$ can be determined and compared to 
 models given in the literature. These form factors are useful for other decays, in particular  B$\to \rho$ + K$^*$ and  B$\to \Phi +{\rm K}^*$.

Motivated by new experimental data\cite{CDF, Babar} and recent theoretical developments\cite {Beneke, Keum}, we are trying  in this paper to investigate some aspects of the B$\to {\rm J}/\Psi + {\rm K}^*$ process.
\section{Decay amplitudes}
\label{section: Decay amplitudes}

{\bf 2-1 Generality, Polarizations and Angular Distributions}

The most general B$\to {\rm V}_1 + {\rm V}_2$ helicity amplitude takes the following form in which we adopt the sign convention of\cite{Dighe}
\begin{equation}
 {\rm H}_{\lambda} \Bigl({\rm B}(P) \to {\rm V}(p_{1},\epsilon_{1}) + {\rm V}(p_{2},
\epsilon_{2})\Bigr)  = \epsilon_1
^{* \mu}(\lambda)
\epsilon_{2}^{* \nu}(\lambda) \left( g_{\mu \nu} {\cal A} +   {P_\mu P_\nu \over m_1 m_2} {\cal B} + \I \epsilon_{\mu\nu\alpha\beta}{p_1^\alpha p_2^\beta \over m_1 m_2} {\cal C} \right) 
\;,
\end{equation}
where $\lambda$ stands for the three helicities $0,\pm 1$ of the massive  
vector mesons with  polarizations $\epsilon_{1}^{\mu}(p_1)$, 
$\epsilon_2^{\nu}(p_2)$. Since the initial B meson is spinless, the two final vector mesons  share the same helicity $\lambda$. Here $M, m_1, m_2$ are masses of the B, V$_1$, V$_2$ mesons  with four-momenta $P, p_1, p_2$  respectively. The ${\cal A}$ and ${\cal B}$  associated to the S and D waves  are  CP-even, while ${\cal C}$ corresponding to the P wave  is CP-odd. We note that in two-body decays, the Lorentz invariant amplitudes ${\rm H}_{\lambda}$  have a mass dimension, so are the quantities 
${\cal A}, {\cal B}, {\cal C}$. From (1), we get
\begin{equation}
 {\rm H}_{0}  = -\Bigl(a\, {\cal A}  + (a^2 -1) \,{\cal B} \Bigr) \;\;\;,\;\;\;\;\;
{\rm H}_{\pm 1}  = {\cal A} \pm  \sqrt{a^2 -1} \, {\cal C} 
\;,
\end{equation}
with\cite{Gourdin} 
\begin{equation}
 a = {p_1 \cdot p_2 \over m_1 m_2} = {M^2-m_1^2-m_2^2\over 2m_1 m_2} \;\;.
\end{equation} 

Also we define K$_c$ as the common momentum  of the outgoing mesons V$_1$ (or V$_2$) in the B rest frame: 
$${\rm K}_c^2 = {\lambda(M^2, m_1^2, m_2^2)\over 4M^2}\;\;,$$
 where $\lambda(x,y,z)= x^2+y^2+z^2-2(xy+xz+yz)$. Thus $\sqrt{a^2 -1}  = M {\rm K}_c / (m_1 m_2)$.

From (1) and (2), the decay rate in each polarization state is given by:
\begin{equation}
  \Gamma_{\lambda} = {{\rm K}_c \over 8 \pi M^2} |H_\lambda|^2
\;,
\end{equation}
 with $\Gamma = \sum_{\lambda} \Gamma_\lambda$ is the full decay width 
$\Gamma ({\rm B}\to {\rm J}/\Psi$ + K$^*$).

 In the following we normalize the partial widths $\tilde{\Gamma}_\lambda = \Gamma_\lambda/\Gamma $ for the three independent polarization states
$$\widetilde{\rm H}_{\lambda} = {{\rm H}_{\lambda} \over \sqrt{
\sum_{\lambda} |{\rm H}_{\lambda}|^2}}\;,$$ 
  such  that $\sum_\lambda |\widetilde{\rm H}_{\lambda}\,|^2  =1$.

The {\it normalized } dimensionless spin amplitudes $A_0$, $A_{\|}$ and $A_{\bot}$ are related to the helicity ones
by
\begin{equation}
 A_0 = \widetilde{\rm H}_{0}\;\;  
 \;,\;\;\; A_{\|} = {\widetilde{\rm H}_{+1} + \widetilde{\rm H}_{-1} \over \sqrt{2}}  \;\;\;,\;\;\; A_{\bot} = {\widetilde{\rm H}_{+1} - \widetilde{\rm H}_{-1} \over \sqrt{2}} 
\;,
\end{equation}
  with again $\sum_\lambda |A_{\lambda}\,|^2  =1$.
To proceed to the $A_0, A_{\|}, A_{\bot}$ determinations, angular measurements are necessary. For that, let us  define the transversity angles $\theta_{\rm tr}$ and $\Phi_{\rm tr}$ as the  polar and azimuthal angles of the $\ell^+$ descended from  ${\rm J}/\Psi \to \ell^+ +\ell^-$ decay in the ${\rm J}/\Psi$ rest frame. The K$^*$ helicity angle $\theta_{K^*}$ is the angle between the K meson direction (coming from  K$^* \to \pi$+K) and the opposite direction of the
${\rm J}/\Psi$ in the K$^*$ rest frame.
The angular distributions\cite{Dighe}  given below  allow us to determine both the  $|\,A_{0}\,|, |\,A_{\|}\,|, |\,A_{\bot}\,|$ magnitudes and phases $\delta_{0}, \delta_{\|}, \delta_{\bot}$ (up to a two-fold 
ambiguity\cite{Suzuki, Bernard}). 
 Thus 
\begin{equation}
  {1\over \Gamma}{\d^3\Gamma\over \d\theta_{\rm tr} \d \theta_{K^*} \d \Phi_{\rm tr}} = {9\over 32 \pi}\Bigl(f_1|A_0|^2 +f_2|A_{\|}|^2 + f_3|A_{\bot}|^2 + 
f_4 Im(A_{\|}^*A_{\bot}) +f_5 Re(A_0^*A_{\|}) +f_6 Im(A_0^*A_{\bot}) \Bigr)
\;,
\end{equation}
where
$$f_1 =2\cos^2 \theta_{K^*} (1- \sin^2\theta_{\rm tr}\cos^2\Phi_{\rm tr}) \;\;,\;\;
f_2 =\sin^2 \theta_{K^*} (1- \sin^2\theta_{\rm tr}\sin^2\Phi_{\rm tr})\;,$$ 
$$f_3 =  \sin^2 \theta_{K^*} \sin^2\theta_{\rm tr} \;\;,\;\;
f_4 = \pm \sin^2 \theta_{K^*}  \sin 2\theta_{\rm tr}\sin\Phi_{\rm tr}\;,$$ 
$$f_5 = - {1 \over \sqrt{2}} \sin 2\theta_{K^*} \sin^2\theta_{\rm tr}\sin 2\Phi_{\rm tr} \;\;,\;\;
f_6 = \pm {1 \over \sqrt{2}} \sin2 \theta_{K^*}  \sin 2\theta_{\rm tr}\cos\Phi_{\rm tr}\;.$$ 
The plus sign in $f_{4,6}$ refers to the B mesons which are $\overline{\rm b}$q$\;$ (q = u, d) bound states, and the minus sign to $\overline {\rm B}$ mesons (b$\overline {\rm q}$ bound states). In the following, for convenience, the amplitudes are implicitly written for the $\overline{\rm B}$ mesons, since we are dealing with the b quark and not the antiquark $\overline{\rm b}$.

 {\bf 2-2 Effective Hamiltonian}

The basis for nonleptonic weak decays of hadrons is the  operator product expansion, and the effective Hamiltonian relevant to B$ \to {\rm J}/\Psi +{\rm K}^*$ may be written as
\begin{equation}
   {\cal H}_{eff} = {G_{\rm F}\over \sqrt{2}} \Bigl( V_{cb} V^*_{cs}\left[C_1(\mu) 
{\cal O}_1(\mu) + C_2(\mu) 
{\cal O}_2(\mu)\right] - V_{tb} V^*_{ts}\sum_{j=3}^6 C_j(\mu){\cal O}_j(\mu)
\Bigr)
\;,
\end{equation}
where the Wilson coefficients $C_i(\mu)$ are evaluated at next-to-leading order and at the renormalization scale $\mu$. We neglect the electroweak penguin operators 
${\cal O}_{7 \cdots 10}$ since their corresponding coefficients $C_7 \cdots C_{10}$ being proportional to $\alpha_{\rm em} =1/137$ are numerically negligible compared to the dominant  $C_1$ and $C_2$ associated to the tree diagrams, and $C_3 \cdots C_6  $ associated to the gluonic penguin loop diagrams. We have:

{\bf V-A current} $\times$ {\bf V-A current}
\begin{equation}
 {\cal O}_1 = (\overline{s}_\alpha c_\alpha)_{V-A} (\overline{c}_\beta b_\beta)_{V-A} \;\;,\;\;
{\cal O}_2 = (\overline{s}_\alpha c_\beta)_{V-A} (\overline{c}_\beta b_\alpha)_{V-A} \;,
\end{equation}
{\bf QCD-Penguins}
\begin{equation}
 {\cal O}_3 = (\overline{s}_\alpha b_\alpha)_{V-A} \sum_{q} (\overline{q}_\beta 
q_\beta)_{V-A} \;\;,\;\;
{\cal O}_5 =(\overline{s}_\alpha b_\alpha)_{V-A} \sum_{q} (\overline{q}_\beta 
q_\beta)_{V+A}\;,
\end{equation}
\begin{equation}
 {\cal O}_4 = (\overline{s}_\alpha b_\beta)_{V-A} \sum_{q} (\overline{q}_\beta 
q_\alpha)_{V-A} \;\;,\;\;
{\cal O}_6 =(\overline{s}_\alpha b_\beta)_{V-A} \sum_{q} (\overline{q}_\beta 
q_\alpha)_{V+A}\;,
\end{equation}
where $\alpha,\beta$ are quark color indices and in (9)-(10), the sum $\sum_q$ is done over  
$q = u, d, s, c, b$ quarks. Also $(\overline{q_1} q_2)_{V - A} (\overline{q_3} q_4)_{V \mp A}$ denotes $ \overline{q_1} \gamma_\rho (1 - \gamma_5) q_2 \; \overline{q_3} \gamma^\rho (1 \mp \gamma_5) q_4$. The coefficients $C_i(\mu)$ are given in Table XXII 
of \cite{Buchalla} at next-to-leading order in the naive dimension regularization (NDR) and in the 't Hooft-Veltman (HV) $\gamma_5$ renormalization  schema :
$$ C_1 = 1.082\; (1.105)\;\;,\;\;C_2 = -0.185 \;(-0.228)\;\;,\;\;C_3= 0.014 \;(0.013) \;,$$ 
$$ C_4 = -0.035 \; (-0.029)\;\;,\;\;C_5 = 0.009 \;(0.009)\;\;,\;\;C_6= -0.041 \;(-0.033)\;,$$
where the first numbers refer to the NDR scheme and those in the parentheses  to the HV scheme, both evaluated at the scale $\mu= \overline{m_b}(m_b) =4.4$GeV. The dependence of $C_i(\mu)$ on the scale $\mu$ as well as  on the regularization-schema must be cancelled in principle by the matrix-elements 
$\bra {\rm K}^* \Psi | {\cal O}_i(\mu)| {\rm B} \ket$ since physical amplitudes $\sim C_i(\mu) \times \bra {\rm K}^* \Psi | {\cal O}_i(\mu)|  {\rm B}\ket$ must be scale and regularization-scheme independent. In the "naive" factorization approach,  $\bra {\rm K}^* \Psi | {\cal O}_i(\mu)| {\rm B}\ket $ is a product of decay constants and form factors, both are real, moreover they are scale and regularization scheme independent, hence the amplitudes via $C_i(\mu)$ suffer from these dependences and turn out to be ambigous. After a tentative approache\cite{Al}, the QCD methods finally solve this problem as we will see.  

  Using the unitarity condition $V_{ub} V^*_{us} +V_{cb} V^*_{cs} +V_{tb} V^*_{ts} =0 $ and neglecting $V_{ub} V^*_{us} \approx 10^{-3}$, we note that (7) has only the unique $V_{cb} V^*_{cs}$ factor. This is crucial to ensure that no matter the penguin $\bra \,{\cal O}_{ 3,6}\,\ket$ and tree $\bra \,{\cal O}_{1,2}\,\ket $ matrix-elements are, we always have
$|\overline{\cal M}/ {\cal M}| =1$ (to a very good precision $V_{ub} V^*_{us} \approx 10^{-3}$), where ${\cal M}$ and $\overline{\cal M}$ are respectively the 
 B$\to {\rm J}/\Psi + {\rm K}({\rm K}^*)$ and $\overline {\rm B} \to {\rm J}/\Psi + \overline{\rm K}(\overline{\rm K}^*)$ amplitudes. This condition  $|\overline{\cal M}/ {\cal M}| =1$ allows us to extract from  experimental data\cite{BetaB, Beta}  the CP asymmetry $\beta$
 angle without any theoretical uncertainties\cite{Hokim} through the "gold-plated" modes B$\to {\rm J}/\Psi + {\rm K}({\rm K}^*)$.

{\bf 2.3 QCD-improved factorization approach}

 To proceed further, we use the QCD-improved factorization approach
\cite{Beneke} according to which, in the  infinite $b$ quark mass limit and for some classes of two-body hadronic B$\to {\rm M}_1 + {\rm M}_2$ decays, the mass-singularities (infrared divergences) {\it factorize}, so that the amplitudes may be written as  convolutions of {\it universal} quantities (the light-cone distribution amplitudes (LCDA)  of the mesons B, M$_1$, M$_2$ with their 
associated semi-leptonic
 form factors $F^{{\rm B}{\rm M}_1}(m_2^2)$) and QCD perturbatively calculable hard-scattering kernels $T^{I}(u), T^{II}(\xi,\eta,u)$ which are 
{\it process-dependent}. 

Here M$_1$ denotes the recoiled meson which can be light ($\pi, \rho$ K, K$^*$ $\cdots$) or heavy (charm D, D$^*$ $\cdots$) but the emitted  M$_2$  can only be a light meson\cite{Beneke} or a Q$\overline {\rm Q}$ quarkonium (heavy or light) and not a heavy meson like D, D$^* \cdots$. The main problem with the latters is that they represent an extended soft hadronic object. In our case M$_1$ is the K$^*$ and M$_2$ is the J$/\Psi$. Schematically we write
\begin{equation}
\bra \overline{\rm K}^* \Psi \, |\;{\cal O}_i\;| \,\overline{\rm B} \ket = 
F^{{\rm B}{\rm K}^*}(m_{\Psi}^2)\int_0^1 \d u T_{\it i}^{I}(u) \Phi_{\Psi}(u) +\int_0^1 \d \xi\, \d \eta \,\d u\, T_{\it i}^{II} (\xi,\eta,u) \Phi_{\rm B}(\xi)\, \Phi_{\rm K^*}(\eta)\, \Phi_{\Psi}(u)
 \;,
\end{equation}
\begin{equation}
\bra \overline{\rm K}^* \Psi \, |\;{\cal O}_i\;| \,\overline{\rm B} \ket = 
\bra   \overline{\rm K}^* \;| {\cal J}_\mu |\, \overline{\rm B}\ket \bra \Psi \;| {\cal J}^\mu | 0 \ket \Bigl[ 1+ \sum r_n \alpha_{s}^n + {\cal O} (\Lambda_{QCD}/ m_b)\Bigr] 
\;,
\end{equation}
where ${\cal O}_i$ are products of different currents ${\cal J}_\mu$
 in (8)-(10).

These equations imply that in the $m_b \to \infty$ limit, "naive" 
factorization (corresponding to the first term 1 in (12)) is recovered in the absence of
 QCD for which $T^{I}(u)$ is independent of $u$ and $T^{II} =0$ in (11). Since the b quark mass is large but finite, power corrections ${\cal O}(\Lambda_{QCD}/ m_b)$ could be significant, specially in some particular cases (for instance in B$ \to \pi +{\rm K}$ for which the scale is not $\Lambda_{QCD}$ but chirally-enhanced like $m^2_{\rm K}/ (m_s +m_d)$).  However it  must be noted  that in the QCD-improved approach\cite{Beneke}, these power corrections (generally associated with nonleading higher-twist LCDA) cannot be reliably computed since in many (but not all) cases, mass singularities  
again show up thus do not 
factorize. 
Keeping this fact in mind, we nevertheless calculate  to order $\alpha_{\rm s}$ the QCD corrections to the naive factorization   B$ \to {\rm J}/\Psi +{\rm K}^*$ amplitude\cite{Gourdin}, the simpler case B$ \to {\rm J}/\Psi +{\rm K}$ of B$\to $V+P has been previously 
studied\cite{Cheng, Chay} within 
the same theoretical framework.

 We also mention  another approach\cite{Keum} called PQCD 
(perturbative QCD) according to which the double logarithms Sudakov suppression effects could regulate the mass singularities, hence power corrections and form factors may  be perturbatively calculable but questioned in\cite{Genon}.

The symbolically written form factors $F^{{\rm B}{\rm K}^*}(m_{\Psi}^2)$ in (11) and  the more explicit  $\bra   \overline{\rm K}^* \;| {\cal J}_\mu |\, \overline{\rm B}\ket$ ones in (12) are in fact defined according to
$$\bra   \overline{\rm K}^*(\epsilon_1, p) \;|\; \overline{s}\gamma_\mu (1-\gamma_5) b\; |\, \overline{\rm B}(P)\ket 
\equiv \bra   \overline{\rm K}^*(\epsilon_1, p) \;|\; V_\mu - A_\mu \; |\, \overline{\rm B}(P)\ket  \equiv {\cal V}_\mu -{\cal A}_\mu \;,$$
where\cite{Bauer}
$${\cal V}_\mu  = \I\;\epsilon_{\mu\nu\alpha\beta}  \;\epsilon_1^{*\nu}(p)\, P^\alpha \,
p^\beta \;{2V(q^2)\over M+ m_{{\rm K}^*}}\;,$$
 $$ {\cal A}_\mu = (M+ m_{{\rm K}^*})  \Bigl[\epsilon^{*}_{1, \mu} -
{\epsilon_1^* \cdot q 
\over q^2} q_\mu \Bigr] A_1(q^2)  -  {\epsilon_1^* \cdot  q 
\over  M + m_{{\rm K}^*}}  
\Bigl[(P+p)_\mu -{M^2-m^2_{{\rm K}^*}  \over q^2} q_\mu \Bigr]  A_2(q^2)$$
\begin{equation}
+ 2m_{{\rm K}^*} A_0(q^2){\epsilon_1^* \cdot q 
\over q^2} q_\mu \;,
\end{equation}
and  $q =P-p $ is the four momentum of the emitted J$/\Psi$ meson.
 In the above equation,  terms proportional to $q_\mu$ vanish when multiplied to the J$/\Psi$ polarization vector $\epsilon_2^{\mu} (q)$.
 With
$$\bra \Psi  \;| {\cal J}^\mu\; | 0 \ket = \bra \Psi (\epsilon_2, q) \;| 
\overline{c} \gamma^\mu \,c \;| 0 \ket = f_{\Psi} m_{\Psi}\,\epsilon_2^{*\mu}(q) $$
where $f_{\Psi} \approx 405$ MeV is the J$/\Psi$ decay constant extracted 
from the J$/\Psi \to \ell^+ +\ell^-$ rate, the first term $\;\bra   \overline{\rm K}^* \;| {\cal J}_\mu |\, \overline{\rm B}\ket \bra \Psi \;| {\cal J}^\mu | 0 \ket$ on the right hand side of (12) is equal to
\begin{equation}
  \epsilon_1^{* \mu}(p)\epsilon_2^{* \nu}(q)  f_{\Psi} m_{\Psi} \Bigl[
-(M+ m_{{\rm K}^*}) A_1(m_{\Psi}^2) g_{\mu\nu} + {2 P_\mu P_\nu \over M + m_{{\rm K}^*}} A_2(m_{\Psi}^2) + {2 \I \,\epsilon_{\mu\nu\alpha\beta} 
 \, p^\alpha \,
q^\beta \over M + m_{{\rm K}^*}}  V(m_{\Psi}^2) \Bigr]\;.  
\end{equation}
 When we compare (1) with (14) using 
(2), (7) and (12), then in terms of an overall common factor
\begin{equation} \kappa \equiv {G_{F}\over \sqrt{2}} V_{cb}V^*_{cs} f_{\Psi} m_{\Psi} (M+ m_{\rm K}^*) \,A_1(m^2_{\Psi})\;\;, 
\end{equation}  
and the dimensionless constants $a, b, c$ given in\cite{Gourdin}
\begin{equation}
  a = {M^2- m^2_{\rm K^*} - m_{\Psi}^2 \over 2 m_{\rm K}^* m_{\Psi} } \; \;,\;\; b= {2 M^2 {\rm K}_c^2 \over m_{\rm K}^* m_{\Psi} (M+m_{\rm K}^*)^2}  
\; \;,\;\; c= {2 M {\rm K}_c \over (M+m_{\rm K}^*)^2} \;,
\end{equation}
 together with the form factor ratios $x, y$  defined by\cite{Gourdin}
\begin{equation}
  x = {A_2(m^2_{\Psi})\over A_1(m^2_{\Psi})  }\; \;,\;\; y= { V(m^2_{\Psi})
\over A_1(m^2_{\Psi}) } \;,
\end{equation}
we obtain the helicity amplitudes H$_{\lambda}$ :  
\begin{equation}
  {\rm H}_{0} =  \kappa \, (a  - b\,x )\,\widetilde{a}_{0} \;\;,\;\;
 {\rm H}_{\pm 1} =  \kappa \, (\pm cy -1) \,\widetilde{a}_{\pm}\;.
\end{equation}
Numerically we have  $a = 3.165, b =
 1.308, c = 0.436$ and the dimensionful factor $\kappa$ (in KeV) is equal to  $2.48 \,A_1(m^2_{\Psi})$.

It remains the most involved computations of the {\it polarization-dependent} coefficients $\widetilde{a}_{\lambda}$ in (18) which are the nonfactorizable $\alpha_{s}$ correction terms  derived from the QCD-improved factorization framework to which the next section will be devoted. In the "naive" factorization method previously considered\cite{Gourdin} without the QCD-improved approach, the $\widetilde{a}_{\lambda}$ was the {\it process-independent} BSW\cite{Bauer} coefficient $a_2= C_2 + C_1/3$, and the penguin
 contributions $C_{3,6}$ are neglected. As is well known, this "naive" factorization method suffers from a serious problem related to scale $\mu$ and regularization-scheme dependences of the Wilson coefficients $C_i(\mu)$, while the amplitudes H$_{\lambda}$ are not. The inclusion of the $\alpha_s$ corrections, as will be shown in (32)-(35), cures this problem.

 For $\overline{\rm B} \to {\rm J}/\Psi + \overline{\rm K}^*$ decay, we remark that the original $V$--$A$ left-handed property of the current 
$\overline{s}\gamma_\mu(1-\gamma_5) b$ in ${\cal O}_i$ is reflected by  the expression (18) of the helicity amplitudes $H_{\pm 1} \sim \pm c V(q^2) - A_1(q^2)$. This may be understood as the following: The $-1/2$ helicity left-handed s quark,  emitted through the $V$--$A$ current, picks up the spectator antiquark 
$\overline{\rm q}$  (which has both $\pm 1/ 2$ helicities) to form the $\overline{\rm K}^*$ meson. Thus the latter can only have $-1$ or $0$ helicity and not $+1$, since the s quark would maintain 
its $-1/2$ helicity  unless final state interactions (FSI) are strong enough to flip it into a $+1/ 2$ state.
Therefore $|{\rm H}_{-1}| \sim |cy +1|$ would largely dominate $|{\rm H}_{+1}| \sim |cy -1|$  unless very strong FSI would reverse the order by making 
$|\widetilde{a}_{+}| \gg |\widetilde{a}_{-}|$. As we will see in section 
{\bf 2-5}, this possibility is not supported by our calculations within the  QCD-improved factorization approach. However the answer as always must come from the experimental side : whether or not 
$|{\rm H}_{-1}| \gg |{\rm H}_{+1}|$ can only be settled by future measurements of the muon polarization in $\overline{\rm B} \to {\rm J}/\Psi + \overline{\rm K}^*$ follows by $  {\rm J}/\Psi \to \mu^+ +\mu^- $.  
 
The decay rate is obtained using (4) and (18) from which we get the normalized longitudinal $A_0$ and transverse $A_{\|}, A_{\bot}$ fractions
 measured by Argus\cite{Argus}, Cleo\cite{Cleo}, CDF\cite{CDF} and 
BaBar\cite{Babar} collaborations
\begin{equation}
|\,A_0\,|^2  = {|\,\widetilde{a}_0\,|^2 (a-bx)^2 \over \Sigma} \;\;,\;\;
\end{equation}
\begin{equation}
|\,A_{\|}\,|^2 =  {|\,\widetilde{a}_{+} +\widetilde{a}_{-}\,|^2 +
 c^2y^2 |\,\widetilde{a}_{+} -\widetilde{a}_{-}\,|^2 - 2 c y \,(|\widetilde{a}_{+}|^2 -|\widetilde{a}_{-}|^2) \over 2\,\Sigma} \;,
\end{equation}
\begin{equation}
 |\,A_{\bot}\,|^2 = {|\,\widetilde{a}_{+} +\widetilde{a}_{-}\,|^2 c^2\,y^2
 +|\,\widetilde{a}_{+} -\widetilde{a}_{-}\,|^2 - 2 c y \,(|\widetilde{a}_{+}|^2 -|\widetilde{a}_{-}|^2) \over 2\,\Sigma} \;\;,\;\;
\end{equation}
where
$$\Sigma=|\,\widetilde{a}_0\,|^2 (a-bx)^2  +  \bigl(|\,\widetilde{a}_{+}\,|^2 +|\,\widetilde{a}_{-}\,|^2 \bigr) (1+ c^2\,y^2) -2 c y\,\bigl(|\,\widetilde{a}_{+}\,|^2 -|\,\widetilde{a}_{-}\,|^2 \bigr)   \;.$$
 The phases $\delta_{0}, \delta_\|, \delta_\bot$ of the $A_{0}, A_{\|}, A_{\bot}$ are given by those of the $\widetilde{a}_{\lambda}$ since
 $\kappa, a, b, c, x, y$ are real.
 
Provided that $|\,\widetilde{a}_{-}\,|^2  \geq |\,\widetilde{a}_{+}\,|^2$ which is true from the first line of (40), and  for ${\it nonzero}$ $A_1(q^2)$, a remarkable upper bound for the longitudinal fraction $\,|A_0\,|^2$ can be derived using (19), no matter how are the finite form factors $A_1(q^2)$, $A_2(q^2) \leq (a/b) A_1(q^2) = 2.42  A_1(q^2)$
 and $V(q^2)$:
\begin{equation}
\,|A_0\,|^2 \leq { a^2\over a^2 + \rho} \;\;,\;\; \rho ={|\,\widetilde{a}_{+}\,|^2 +|\,\widetilde{a}_{-}\,|^2  \over |\,\widetilde{a}_0\,|^2 }\;,
\end{equation}
the derivation of this upper bound can be easily obtained by considering the lower bound of the inverse $1 / |A_0\,|^2$.
Of course when all of the three  $\widetilde{a}_{\lambda}$ are real and identical, we recover our old result\cite{Gourdin} of the naive factorization method, as it should be:
\begin{equation}
|\,A_0\,|^2  = { (a-bx)^2 \over (a-bx)^2 + 2 (1 + c^2 y^2)} \leq {a^2\over a^2 +2} = 0.83 \;\;,\;\;
\end{equation}
\begin{equation}
 |\,A_{\|}\,|^2 = { 2  
\over (a-bx)^2 +2 (1 + c^2y^2)} \;,
\end{equation}
\begin{equation}
 |\,A_{\bot}\,|^2  = { 2 c^2y^2 
\over (a-bx)^2 +2 (1 + c^2y^2)} \;.
\end{equation}
The latest experimental data for $A_{0}, A_{\|}, A_{\bot}$ are\cite{CDF, Babar}
 $$|\,A_{0}\,|^2 = 0.597 \pm 0.028 \pm 0.008 \;\;,\;\;
 |\,A_{\bot}\,|^2 = 0.160 \pm 0.032 \pm 0.036 \;\;,\;\;$$
\begin{equation} 
|\,A_{\|}\,|^2 = 1- |\,A_{0}\,|^2 -|\,A_{\bot}\,|^2  = 0.243 \pm 0.034 \pm 0.033\;.
\end{equation} 
From the experimental value of $|\,A_0\,|^2 =0.6$, we derive a constraint on the ratio 
$\rho$ using (22),
$${|\,\widetilde{a}_{+}\,|^2 +|\,\widetilde{a}_{-}\,|^2  \over |\,\widetilde{a}_0\,|^2 } \leq {2\over 3} a^2 = 6.6$$
 which is  however far from being saturated (section {\bf 2-5} below).

For the phases $\delta_i$, since measurements of the interference terms in the angular distributions are limited to Re$(\,A_{\|} A^*_{0})$, Im$(\,A_{\bot} A^*_{0})$ and Im$(A_{\bot} A^*_{\|})$, there exists a two-fold ambiguity\cite{Suzuki, Bernard}  
$$\delta_{\|} \leftrightarrow - \delta_{\|} , \delta_{\bot} \leftrightarrow  \pi- \delta_{\bot} , \delta_{\bot} - \delta_{\|} \leftrightarrow \pi - (\delta_{\bot} - \delta_{\|})\;.$$
 The phases quoted in radians are\cite{Babar,Suzuki}
 $$\delta_{\bot} \equiv {\rm arg}(\,A_{\bot} A^*_{0}) = -0.17 \pm 0.16 \pm 0.06 \;(-2.97  \pm 0.16 \pm 0.07) \;\;,\;\;$$
\begin{equation} 
\delta_{\|} \equiv {\rm arg}(\,A_{\|} A^*_{0}) = 2.50 \pm 0.20 \pm 0.07 \; (- 2.50 \pm 0.20 \pm 0.07 )\;,
\end{equation} 
the numbers in parentheses correspond to the second solution due to the mentioned ambiguity.
Consequently, from (26)-(27), one deduces\cite{Suzuki} either $|{\rm H}_{+1}/{\rm H}_{-1}| =0.26 \pm 0.14$ or  $|{\rm H}_{-1}/{\rm H}_{+1}| =0.26 \pm 0.14$, this ambiguity can be solved in the future by the J$/\Psi \to \ell^+ +\ell^-$ lepton polarization 
measurements. 

The most important information we can draw from the measured nonzero phase of
 $\delta_{\|}$ is that nonfactorizable corrections to the  "naive" factorization method 
must be taken into account.

{\bf 2-4 Nonfactorizable Corrections} 

In the QCD-improved factorization approach, the light-cone distribution amplitudes  play a central role. For vector mesons, the LCDA are given by\cite{Braun, Feldmann}
$$
\bra V(p, \epsilon)\;|\overline{q}_i(y) q_j(x) \;| 0\,\ket = {1\over 4} \int_0^1 \d \eta \;e^{\I(\eta p\cdot y + (1-\eta)p\cdot x)}\,\Bigl(f_{V} M_{V} 
\Bigl[\Esl_{\|} \Phi_{\|}(\eta) + \Esl_{\bot} \varphi^v_{\bot}(\eta)\Bigr]_{ij}$$
\begin{equation} 
+ F_{V}^{T} \Bigl(\Esl_{\bot} \psl_{\bot}\Bigr)_{ij} \Phi_{\bot}(\eta) +
 \Bigl[ 1- {2m_q F_{V}^{T} \over f_{V} M_{V} } \Bigr] \varepsilon_{\mu\nu\alpha\beta}(\gamma^\mu \gamma_5)_{i j} \epsilon^{* \nu} p^{\alpha} z^{\beta} \varphi^{\rm a}_{\bot}(\eta)   \Bigr)
\end{equation}
where $i, j$ denote the Dirac spinor indices, $z  = y-x$, 
$\epsilon^{\mu}_{\|}$ ($\epsilon^{\mu}_{\bot}$) are the longitudinal (transverse) polarizations of the vector meson and $f_V$ and $F^T_V$ 
are respectively   its vector and tensor decay constants defined by
$$\bra V(p, \epsilon)\,|\overline{q}(x)\gamma_\mu q(0)\,| \,0 \ket = f_V M_V {\epsilon \cdot x \over p \cdot x} p_\mu \int_0^1 \d \eta \; e^{\I \eta p \cdot x} \Phi_{\|}(\eta) \;, $$
\begin{equation} 
\bra V(p, \epsilon)\,|\overline{q}(x)\sigma_{\mu \nu} q(0)\,| \,0 \ket = 
-\I \,F_{\it V}^{\it T}  (\epsilon_\mu  p_\nu - \epsilon_\nu  p_\mu )  \int_0^1 \d \eta \;e^{\I \eta p\cdot x} \Phi_{\bot}(\eta) \;.
\end{equation}
Contracting the above equation $\bra V(p, \epsilon)\,|\overline{q}(x)\sigma_{\mu \nu} q(0)\,| \,0 \ket$ with $p^\nu$ and applying the equation of motion
 together with the definition $\bra V(p, \epsilon)\,|\overline{q}(0)\gamma_{\mu } q(0)\,| \,0 \ket = f_V M_V \epsilon_\mu $, a relation is obtained
 between the $f_V$ and $F^T_V$ decay constants\cite{Cheng, Chay}:
\begin{equation}
F^T_V \, M_V = 2 f_V \, m_q  \;\;
\Longrightarrow {F^T_V m_q\over f_V M_V} =2 \left({m_q\over M_V}\right)^2 = 2 \eta^2 \;,
\end{equation}
where in the last step the on-shell relation 
$\eta^2 \psl^2 =m_q^2$  has been applied\cite{Cheng}. 

Finally 
 $\Phi_{\|}(\eta)$, $\Phi_{\bot}(\eta)$ are the leading twist-2 LCDA amplitudes while the 
nonleading vector-like twist-3 LCDA is denoted by $\varphi^{\rm v}_{\bot}(\eta)$. The axial-like twist-3 LCDA $\varphi^{\rm a}_{\bot}(\eta)$ contribution  is negligible since it is proportional to 
$f_V M_V - 2 F^T_V m_q = M_V^2-4 m_q^2 \approx 0$.  We thus introduce expressions $\Phi^{{\rm K}^*}_{\|}(\eta)$, $\Phi^{{\rm K}^*}_{\bot}(\eta)$, $\varphi^{{\rm K}^* {\rm v}}_{\bot}(\eta)$ and
$\varphi^{{\rm K}^* {\rm a}}_{\bot}(\eta)$ for the vector meson K$^*$.

 For the pseudoscalar B meson, its wave function in the heavy b quark limit is 
\begin{equation}
 \bra \,{\rm B}(P)\,| \overline {b}_i(x)  q_j(0) \,| 0 \ket|_{x_{+} = x_{\bot} =0}  = {\I \,f_{\it B} \over 4} \int_0^1 \d \xi \; e^{\I \xi \, P_{+} \cdot x_{-}} \Bigl[ (\Psl + M) \gamma_5 \Phi_{\rm B}(\xi) +\cdots \Bigr]_{i j}\;, 
\end{equation}
where $\cdots$ denote terms that do not contribute to the decay amplitude calculated later in (39)--(40).

For the vector meson J/$\Psi$ treated as a heavy charmonium, to the leading order in $1/m_{\rm c}$, its wave function has a similar expression:
$$ \bra \,{\rm J}/\Psi(p,\epsilon)\,| \overline {c}_i(x)  c_j(0) \,| 0 \ket|_{x_{+} = x_{\bot} =0}  = {f_{\Psi} \over 4} [\Esl (\psl + m_{\Psi} )]_{j k} \int_0^1 \d u \; e^{\I u  p_{+} \cdot x_{-}} \Bigl[ \Phi_{\psi}(u) +\cdots \Bigr]_{k i}
\;. $$

We note that  the use of the light-cone wave function for the heavy J/$\Psi$ is problematic, higher twist effects have to be included, however they may not converge fast enough. We adopt in the following $\Phi_{\|}(u)$ as the distribution amplitude (DA) of the nonlocal vector current of J/$\Psi$, thus we treat the  J/$\Psi$ wave function on the same footing as the B meson. Comparing the above equation with  (28), we see that at the leading order in $1/m_{\rm c}$ one has for the heavy charmonium J$/\Psi$ :
$$\Phi_{\|}(u) = \Phi_{\bot}(u) = \Phi_{\psi}(u)\; \;{\rm and} \; {\rm F}^T_{\Psi} = f_{\Psi} \;.$$
Equipped with these ingredients, we are ready to compute the nonfactorizable correction terms. 

Loop integrations of the four vertex corrections diagrams 
(Fig.6 in\cite{Beneke} which gives $F_I^\lambda$) and the two spectator diagrams (Fig.8 
in\cite{Beneke} which gives $F_{II}^\lambda$) are not detailed here. Only we give the results in the NDR scheme:
\begin{equation}
\widetilde{a}_{\lambda} =(a_2^{\lambda} + a_3^{\lambda}+ a_5^{\lambda})\;,
\end{equation}
\begin{equation}
a_2^{\lambda} = C_2 + {C_1 \over N_c } +{\alpha_s \over 4 \pi} { N_c^2 -1 
\over 2 N_c^2 } C_1 \Bigl(-18 +12 \ln {m_b \over \mu} + 
F_I^{\lambda} + F_{II}^{\lambda} \Bigr)\;,
\end{equation}
\begin{equation}
a_3^{\lambda} = C_3 + {C_4 \over N_c } +{\alpha_s \over 4 \pi} { N_c^2 -1 
\over 2 N_c^2 } C_4 \Bigl(-18 +12 \ln {m_b \over \mu} + 
F_I^{\lambda} + F_{II}^{\lambda} \Bigr)\;,
\end{equation}
\begin{equation}
a_5^{\lambda} = C_5 + {C_6 \over N_c } - {\alpha_s \over 4 \pi} { N_c^2 -1 
\over 2 N_c^2 } C_6 \Bigl(-6 +12 \ln {m_b \over \mu} + 
F_I^{\lambda} + F_{II}^{\lambda} \Bigr)\;,
\end{equation}
 with $N_c=3$ is the color number.
 For completeness, the constants $-18, -18, -6$ in (33)-(35) become 
respectively $-14, -14, -18$
in the HV scheme. These constants and the $\ln(m_b/\mu)$  term inside the parentheses of (33)-(35)  reflect the  scale and regularization-scheme dependences, they are cancelled by those 
of the Wilson coefficients $C_i(\mu)$, therefore the final expressions of 
$\widetilde{a}_{\lambda}$ are $\mu$ and regularization-scheme independent.
The $F_I^\lambda$ and $F_{II}^\lambda$ are calculated to be:  
$$F_I^0 =  \int_0^1 \d u\Bigl[ \Phi_{\|}(u) \,
{\cal K}(r,u) + \Phi_{\bot}(u) {\cal H}(r,u) \Bigr]\;$$ 
where
$$
{\cal K}(r,u) = 3 \,[\;\ln(1-r) - \I \pi]  + { 3(1-2u) \ln u \over 1-u}  + {2 {\it r} (1-u)\over 1-{\it r} u} 
$$
\begin{equation}
 + \Bigl[ {1-u\over (1-r u)^2} -{u\over [1-r(1-u)]^2} \Bigr] r^2 u \ln (r u) + {r^2 u^2 [\;\ln (1-r) -\I \pi] \over [1-{\it r} (1-u)]^2} \;,
\end{equation}
and
\begin{equation}
{\cal H}(r,u) =  8r \,u^2 \Bigl\{ \Bigl[ {1\over 1-{\it r} (1-u)} -{ 1\over 1- {\it r} u}\Bigr] \ln({\it r} u) -{\ln(1-{\it r}) -\I \pi \over 1-{\it r} (1-u)} \Bigr\} \;,
\end{equation}
with $r= m^2_{\Psi}/M^2$. Also
$$
F_I^{\pm} = \int_0^1 \d u\; \varphi^v_{\bot}(u)\,
{\cal K}(r,u) $$
$$+ 8\int_0^1 \d u \, \Phi_{\bot}(u) u^2 \Bigl \{ -{ln u \over 1-u} +{{\it r} 
\ln ({\it r} u) \over 1- {\it r}(1-u) } - {\it r}\,  {\ln (1-{\it r}) -\I \pi \over 1-{\it r} (1-u)} 
\Bigr \}$$
$$
+\int_0^1 \d u \Phi_{\|}(u) \Biggl \{  
{(3-2u)\ln u \over 1-u} + { 2{\it r} u \over 1-{\it r}(1-u)} + \Bigl (- {3 \over 1-{\it r}u} + {1- {\it r}(1+u)\over [1-{\it r}(1-u)]^2}
\Bigr ) {\it r}u \ln ({\it r} u) $$
\begin{equation}
 + \Bigl(3 - r(6-5u) + r^2 (3-5u + 4u^2) \Bigr){\ln(1-r) -\I \pi \over [1-{\it r}(1-u)]^2} \Biggr \} \;.
\end{equation} 

Similarly to the B$\to J/\Psi$ +K case found in\cite{Cheng, Chay}, the infrared divergences in $F_I^{\lambda}$ are mutually cancelled among the four vertex diagrams, this cancellation is  essentially due to the symmetric 
$u \leftrightarrow 1-u$ of 
the kernels $
{\cal K}(r,u)$,  $
{\cal H}(r,u)$ and the wave  functions.
 We note that in (38), the J/$\Psi$ nonleading $\varphi_{\bot}^{\rm v}(u)$ wave function also contributes  to the transverse $F_I^{\pm}$ on the same footing as the leading $\Phi_{\bot}(u)$.

For the $F_{II}^{\lambda}$ of the spectator-quark effect, following\cite{Cheng2} we get
\begin{equation}
F_{II}^{0} ={4\pi^2\over N_c} {f_B f_{{\rm K}^*}  \over  m_{\Psi} 
(M+m_{{\rm K}^*}) A_1(m^2_{\Psi}) (a-bx)}  \int_0^1 \d \xi {\Phi_{\rm B}(\xi)\over \xi} \int_0^1 \d \eta {\Phi^{K^*}_{\|}(\eta)\over \eta}\int_0^1 \d u {\Phi_{\|}(u)\over u}
\;,
\end{equation}
$$
F_{II}^{\pm} = {16 \pi^2 \over N_c} {f_B f_{\Psi} (1\mp 1) \over M (M+m_{{\rm K}^*})
 A_1(m^2_{\Psi})( 1 \mp cy) }
\int_0^1 \d \xi {\Phi_{\rm B}(\xi)\over \xi} \int_0^1 \d \eta 
{\Phi^{K^*}_{\bot}(\eta)\over \eta^2} \int_0^1 \d u \, \varphi^v_{\bot}(u)\; u   $$
$$- {8\pi^2\over N_c} {f_B f_{{\rm K}^*} m_{{\rm K}^*} \over M^2 
(M+m_{{\rm K}^*}) A_1(m^2_{\Psi})(1 \mp cy) }   
 \int_0^1 \d \xi {\Phi_{\rm B}(\xi)\over \xi} \int_0^1 \d \eta \d u
\Bigl\{ \varphi^{K^*  v}_{\bot}(\eta) \varphi^{v}_{\bot}(u) {\eta +u \over \eta^2 u} $$
\begin{equation}
\pm {1\over 4} \varphi^{K^* {\rm v}}_{\bot}(\eta)\varphi^{\rm a}_{\bot}(u) {\eta +u \over \eta^2 u^2}
\mp {1\over 4} \varphi^{K^* {\rm a}}_{\bot}(\eta)\varphi^{\rm v}_{\bot}(u) {\eta + 2u  \over \eta^3 u} \Bigr\}
\;.
\end{equation}
We first emphasize that the infrared-finite longitudinal $F_{II}^{0}$ in (39) is {\it not} $1/ M$ power-suppressed contrary to its appearance, since the B meson wave function 
$\Phi_{\rm B}(\xi)$ is appreciable only for $\xi$ of the order $\Lambda_{QCD}/ M$, hence the integral $\int \d \xi \Phi_{\rm B}(\xi)/ \xi \sim M/\Lambda_{QCD}$ compensates the $f_{\rm K^*}/(M+m_{{\rm K}^*})$ in (39).

As for the transverse  $
F_{II}^{\pm} $ parts given in (40), they  are {\it infrared divergent} although
  $1/M$ and $1/M^2$ power-suppressed. Indeed, the first term in $F_{II}^{-} \sim \int \Phi^{{\rm K}^*}_{\bot}(\eta)/ \eta^2$ unexpectedly diverges {\it even with} the K$^*$  {\it leading}  twist-2 LCDA $\Phi^{{\rm K}^*}_{\bot}(\eta)$, the  remaining divergent terms come from the twist-3 LCDA of both the K$^*$ and J$/\Psi$ mesons. We have neglected the $r$ dependences in  
$F_{II}^{\pm}$ to simplify the computations of complicated loop integrals, since infrared divergences occur no matter the $r$ dependences are kept or not. 

In the numerical applications, we use for the leading twist-2 LCDA their asymptotic form
$\Phi_{\|}(x) = \Phi_{\bot}(x) = 6x(1-x)$ for both J$/\Psi$ and K$^*$ mesons, although the former is treated as heavy. The twist-3 LCDA of the K$^*$ vector meson are taken to be $\varphi^{\rm v}_{\bot}(x) =  3[1+(2x-1)^2]/4$ and $\varphi^{\rm a}_{\bot}(x)= 6x(1-x)$. For the B meson wave function, we take $\Phi_{\rm B}(\xi) = N_B \xi^2(1-\xi)^2 \exp[- \xi^2 M^2 /2 \omega^2]$ with $\omega =0.25 $ GeV, and $N_B$ is the normalization factor such that $\int_0^1  \d \xi \,\Phi_{\rm B}(\xi) =1$. With this $\Phi_{\rm B}(\xi)$, we get $\int_0^1  \d \xi \,\Phi_{\rm B}(\xi)/ \xi =
M/ (0.3 {\rm GeV})$, in agreement with our guess  $\int \d \xi \Phi_{\rm B}(\xi)/ \xi \sim M/\Lambda_{QCD}$ mentioned above. 

{\bf Remarks}

From (36)-(40) we draw some unexpected features of the nonfactorizable terms in B$\to {\rm J}/\Psi +{\rm K}^*$ which are dictinctive from  
B$\to {\rm J}/\Psi +{\rm K}$ :

1- The chirally-enhanced factor $ m_{\rm K}^2/ (m_s+m_d)$ inherent to the pseudoscalar K meson in B$\to {\rm J}/\Psi +{\rm K}$ is absent in 
B$\to {\rm J}/\Psi +{\rm K}^*$ with the vector ${\rm K}^*$ meson. Therefore, neither $F_I^{\lambda}$ nor $F_{II}^{\lambda}$  are chirally-enhanced in   both the longitudinal and transverse parts.

 2- Both the $1/m_{\rm c}$ leading and nonleading wave functions of the charmonium J$/\Psi$ contribute to the transverse $F_I^{\pm}$ and $F_{II}^{\pm}$, moreover $F_I^{\pm}$ is {\it infrared-finite} with the nonleading $\varphi^{\rm v}_{\bot}(u)$ as shown in the first line of (38). The longitudinal 
$F_I^{0}$ and $F_{II}^{0}$  are also {\it infrared-finite} 
and {\it not} power-suppressed.

3- At the first order $\Lambda_{QCD}/M$ level, $F_{II}^{+}$ vanishes, only survives the infrared-divergent $F_{II}^{-}$. Unexpectedly, the infrared divergence of $F_{II}^{-}$ already comes from the leading twist-2 LCDA of the K$^*$, its twist-3 is unnecessary to render  $F_{II}^{-}$  divergent. At the second order 
$\Lambda^2_{QCD}/M^2$, the  twist-3 $\varphi^{\rm K^*}_{\bot}(\eta)$ and
$\varphi_{\bot}(u)$ of both K$^*$ and J$/\Psi$ respectively make $F_{II}^{\pm}$ infrared divergent. 

4- Fortunately, the longitudinal part $\widetilde{a}_0$ as given by (36) and 
(39) is infrared convergent, therefore it can be unambigouly used  in the following section to check whether or not agreement exists between 
theoretical calculations and 
experimental data.

{\bf 2-5 Numerical Results}

In  (39) we need the decay constant $f_{\rm K^*} $ of the vector meson K$^*$, 
for that  we may use the $\tau$ lepton decay $\tau \to \nu_{\tau} +{\rm K^*}$ width given by
$$\Gamma(\tau \to \nu_{\tau} +{\rm K^*}) = {G_F^2 |V_{us}|^2 f^2_{\rm K^*} M_{\tau}^3\over 8\pi} \Bigl(1+ {2 m^2_{\rm K^*} \over  M_{\tau}^2} \Bigr)\Bigl
(1- { m^2_{\rm K^*} \over  M_{\tau}^2} \Bigr)^2 \;,$$ 
to extract $f_{\rm K^*} \approx  210$ MeV, a value consistent with $ m_{\rm K^*}f_{\rm K^*} = m_{\rho} f_{\rho}$ of 
the SU(3) flavor symmetry which relates $f_{\rm K^*} $ to the 
decay constant $f_{\rho} \approx 198$ MeV of the charged $\rho(770)$ vector meson. 
 We also take
 $f_{\Psi} \approx  405$ MeV and $f_{\rm B} \approx  180$ MeV. 

Numerical values of the infrared-finite quantities $F_I^0$ in (36)-(37) and  
$F_{II}^0$ in (39) are
\begin{equation}
F_I^0 = -0.82 - \I \;6.61 \;\;,\;\;
 F_{II}^0 ={4.72\over A_1(m^2_{\Psi} ) (a-bx)}\;.
\end{equation}
These numerical values may vary within $20$ per-cent when different LCDA are used instead of their asymptotic forms.
Since both $F_I^0$ and $F_{II}^0$ are finite, we  first concentrate on the longitudinal part  $$\Gamma_{\rm L}({\rm B} \to J/\Psi +{\rm K^*}) = |A_0|^2 {{\rm B}{\rm r}({\rm B} \to J/\Psi +{\rm K^*}) \over \tau_{\rm B}}$$ where Br$({\rm B} \to J/\Psi +{\rm K^*})$ is the branching ratio $= (1.35 \pm 0.18)\times 10^{-3}$ for the neutral and $ (1.47 \pm 0.27)\times 10^{-3}$ for the charged B meson and $\tau_{\rm B}$ is their respective lifetime $[1.56 (1.65) \pm 0.04] \times 10^{-12}\; {\rm s}$. We obtain on average  $\Gamma_{\rm L}({\rm B} \to J/\Psi +{\rm K^*}) = (3.37 \pm 0.4)\times 10^{-16}$ GeV, using $|A_0|^2 =0.6$. When this experimental value is compared  with the theoretical expression
$$\Gamma_{\rm L}({\rm B} \to J/\Psi +{\rm K^*}) ={{\rm K}_c \over 8\pi M^2} \kappa^2 (a-bx)^2 |\widetilde{a}_0|^2\;,$$ we find that the product $ (a-bx) A_1(m_{\Psi}^2) |\widetilde{a}_0|$ is constraint to equal  $0.156 \pm 0.02$, thus
\begin{equation}  (a-bx) A_1(m_{\Psi}^2) |\widetilde{a}_0| = 0.156 \pm 0.02 \;.
\end{equation}
 Our formulae (33)--(39) with $\alpha_s(m_b) = 0.23$ give $|\widetilde{a}_0| \approx 0.14$. 

This value of $|\widetilde{a}_0|$ in turn can  easily satisfy the constraint (42), and we get a domain for $x, A_1(m_{\Psi}^2)$ plotted by 
a  hyperbolic curve and translated into the following numerical values:
\begin{equation}
0 < x \leq 1.1  \;\;\;\; \,\,\, {\rm and}\;\;\;\; 0.35 < A_1(m_{\Psi}^2) \leq 0.60 \;.
\end{equation}
The smallest x is associated with the smallest  $A_1(m_{\Psi}^2)$, the latter increases with increasing $x$. 
For the  transverse $\widetilde{a}_{\pm}$ which cannot be reliably calculable because of 
their infrared divergences, we  reverse the naive factorization procedure previously proposed\cite{Gourdin} in which (23) was used to determine $x, y$. Now we fix 
$x \approx 1.1$ and $|\widetilde{a}_0| =0.14$ then using the theoretical expressions (19)-(21) matched with the experimental data (26)-(27), we determine in turn $\widetilde{a}_{\pm}$ and $y$. The resulting contour solutions $|\widetilde{a}_{+}| \approx 0.095 \pm 0.02$, $|\widetilde{a}_{-}| \approx 0.125 \pm 0.02$  confirm the polarization-dependence of $\widetilde{a}_{\lambda}$. Also we get $y \approx 1.75$.

We remark that our favoured  values 
$x\equiv A_2(m_{\Psi}^2)/A_1(m_{\Psi}^2) \leq 1.1$ and $y\equiv V(m_{\Psi}^2)/A_1(m_{\Psi}^2)\approx 1.75$ are generally satisfied by some models of form factors studied in the literature\cite{Ball, Ali, Khod, Charles, Melikhov}. It is amusing to note however that $x\leq 1.1$ is at variance with the  B-K$^*$ form-factor ratio derived below from equation of motion for on-shell {\it massless} strange quark. Indeed using
$$p^\mu \overline{s}\gamma_\mu \gamma_5 b\ = m_s \overline{s} \gamma_5 b =0 
\;,$$  then we obtain a relation between the two form factors $A_1(q^2)$ and 
$A_2(q^2)$. Assuming 
\begin{equation}  
 p^\mu \bra  \overline{\rm K}^*(\epsilon_1, p) \;|\; \overline{s}\gamma_\mu \gamma_5 b\; |\, \overline{\rm B}(P)\ket  \equiv p^\mu {\cal A}_\mu  = 0\;,
\end{equation}
which gives
$$
 A_1(q^2) = {\lambda(M^2, m^2, q^2)\over (M+m)^2 (M^2-m^2-q^2)} A_2(q^2) 
+{2m \over M+m} A_0(q^2)\;,$$
we get
\begin{equation} 
{A_2(q^2)\over A_1(q^2) } = { (M+m)^2 (M^2-m^2-q^2)\over  \lambda(M^2, m^2, q^2)} \Bigl[ 1- {2m \over M+m} {A_0(q^2)\over A_1(q^2) }\Bigr]\;.
\end{equation}

From (45), we recover the well-known relation at $q^2=0$
$${M+m\over 2m}  A_1(0) - {M-m\over 2m} A_2(0) = A_0(0)\;.$$ 
Neglecting $m^2 $ with respect to $ M^2$ and $ q^2 = m^2_{\Psi}$, then we get from (45)
$$ x \equiv { A_2(m_{\Psi}^2) \over A_1(m_{\Psi}^2)} ={M^2\over
 M^2-m_{\Psi}^2} = 1.52 \;,$$
this too large value  indicates that  $m_s =0$ in (44) 
may not be  a good approximation.

Although the relation (45) is derived here with the assumption  $m_s =0$,  we  remark nevertheless that it bears some similarity with the one\cite{Charles} derived in the very different context of the  large recoil energy $q^2$ 
limit for which $m$ should be neglected:
$$ 
{A_2(q^2)\over A_1(q^2) } = {(M+m)^2 \over M^2 -m^2-q^2} \Bigl[1 - {2Mm\over  M^2 -m^2-q^2} {\xi_{\|} \over  \xi_{\bot}}\Bigr] \Rightarrow {M^2\over M^2-q^2}.$$  
 This is only this large $q^2$ recoil energy limit that can justify the  above result of\cite{Charles} and not the $m_s =0$ assumed here in (44).

\section{Conclusion}
We have examined within the QCD-improved factorization approach different aspects of the color-suppressed B decay into two vector mesons 
B$\to J/\Psi + K^*$ for which important experimental results are recently 
obtained\cite{CDF, Babar}. The nonzero phases of 
the helicity amplitudes  measured  for the first time by\cite{CDF, Babar} 
indicate that nonfactorizable terms must be taken into account. 
We emphasize that the phases can only be determined in B$\to$ V + V decay, 
hence its superiority over the B$\to$ V + P and  B$\to$ P + P in this aspect. Our calculations (36)-(39)  give  nonzero imaginary part to the $process$-$dependent$ and 
$polarization$-$dependent$ coefficients $\widetilde{a}_\lambda (\Psi {\rm K}^*)$ that substitute  the conventional  process-independent and 
polarization-independent  BSW $a_2$ coefficient. The phases obtained 
for our $\widetilde{a}_{\lambda} (\Psi {\rm K}^*)$  are consistent with experiments.  

We find that the longitudinal $\widetilde{a}_0 (\Psi  {\rm K}^*)$ is free of infrared 
divergence and  $|\widetilde{a}_0 (\Psi  {\rm K}^*)| \approx 0.14$ is about twice the  BSW
$a_2 \approx 0.074$, thus  corrections -- mainly due to the spectator-quark effect  $F_{II}^0$ in (38) -- are large but under control. This $|\widetilde{a}_0 (\Psi {\rm K}^*)|$ is also different from the one in B$\to$J$/\Psi $ +K case for which experimental data indicate 
that $|\overline{a_2} (\Psi  {\rm K})| \approx 0.25$, 
thus confirming their process-dependence. 

On the other hand, our calculations show that the transverse part $\widetilde{a}_{\pm} (\Psi {\rm K}^*)$ is infrared divergent 
(although power-suppressed), this infrared divergence may be handled by a cutoff procedure. From  remarks in ${\bf 2.4}$, we note an important difference between 
B$\to {\rm J}/\Psi +{\rm K}$ and B$\to {\rm J}/\Psi +{\rm K}^*$  in their nonfactorizable terms. In the former case, the discrepancy by a factor of three between experimental data ($|\overline{a_2} (\Psi  {\rm K})| \approx 0.25$) and the theoretical estimates\cite{Cheng, Chay}  
($|\overline{a_2}(\Psi  {\rm K})| \approx 0.08$) using twist-2 LCDA may be cured\cite{Cheng} by introducing the chirally-enhanced  twist-3 LCDA of the K meson which gives a formally infrared divergent
$\overline{a_2}(\Psi  {\rm K})$. This divergence is parameterized\cite{Beneke, Cheng} by a random number $X$  just to render $\overline{a_2}(\Psi  {\rm K})$ finite and {\it large enough} to fit data. This input cannot be evoked here for the B$\to {\rm J}/\Psi +{\rm K}^*$ case, since the longitudinal $\widetilde{a}_{0} (\Psi {\rm K}^*)$ (the equivalence of $\overline{a_2}(\Psi  {\rm K})$) is {\it finite}. Moreover, if one postulates that the input $X$ is used to cure the infrared divergence of $\widetilde{a}_{\pm}(\Psi {\rm K}^*)$ -- similarly to the B$\to {\rm J}/\Psi +{\rm K}$ case where $X$ cures
 the  $\overline{a_2}(\Psi  {\rm K})$ -- then the experimental data on $|A_{\|}|, 
|A_{\bot}| < |A_0|$ imply that the input $X$ is constraint to make the divergent $\widetilde{a}_{\pm}(\Psi {\rm K}^*)$ be smaller than 
the convergent $\widetilde{a}_{0}(\Psi {\rm K}^*)$, 
which is somewhat disturbing.

Therefore we are inclined to believe that in our case of B$\to $J/$\Psi +{\rm K}^*$, the procedure used to handle the infrared divergence via $X$  may be  inadequate for the treatment of the discrepancy (if any) between experimental data and  theoretical  estimates using QCD-improved factorization approach. We may seek the remedy outside the $\widetilde{a}_{\lambda}$, probably in the form factor $A_1(m_{\Psi}^2)$ and in the ratios $x =A_2 / A_1$ and $y =V/A_1$, since the overall factor $A_1(m_{\Psi}^2)$ in (15) is {\it central} to the absolute strength of the decay rate B$\to {\rm J}/\Psi +{\rm K}^*$, as well  the ratio $x$ is {\it central} to the longitudinal part $A_0$ and $y$ to the transverse fraction $A_{\bot}$. To deal with the  transverse part $\widetilde{a}_{\pm} (\Psi {\rm K}^*)$, we  adopt a pragmatic method by fixing $x$ and $|\widetilde{a}_0 (\Psi  {\rm K}^*)|$ previously obtained from $|A_0|$, then  using data on $|A_{\|}|$ and $|A_{\bot}|$ together with their theoretical expressions (19)-(21), we determine in turn 
 $|\widetilde{a}_{+} (\Psi  {\rm K}^*)| \approx 0.095 \pm 0.02$, $|\widetilde{a}_{-} (\Psi  {\rm K}^*)| \approx 0.125 \pm 0.02$. Moreover the ratio $y \equiv V(m_{\Psi}^2) 
/A_1(m_{\Psi}^2)$ is also bounded around $1.75$. The constraints  $x \leq 1.1$ and $y \approx 1.75$ have implications on models of B-K$^*$ form factors commonly used in the literature. 

In summary, our results show that the spectator effects and final state interactions reflected by $F^0_{II}$  play an important role 
in our  quantitative understanding of the color-suppressed B$\to$ J$/\Psi$ +K$^*$ decay for the dominant longitudinal mode. However the power $\Lambda_{QCD}/m_b$ corrections  within the QCD-improved factorization approach has to be better understood.

{\it Note added}: After the first version of this paper is circulated, we received a preprint hep-ph/0111094 by H.Y. Cheng, Y.Y. Keum and K.C. Yang. The second version  takes into account
 their greatly appreciated remarks. We thank the referee of JHEP for her (his) critical  remarks, helping us to improve the paper in its third version. 

\begin{em}

\end{em}
\end{document}